\documentclass[letterpaper,reprint,prl,superscriptaddress,showpacs,longbibliography,twocolumn]{revtex4-2}
\usepackage{xcolor}
\usepackage{mathrsfs}
\usepackage{amsmath}
\usepackage{amssymb}
\usepackage{graphicx}
\usepackage{wasysym}
\usepackage{bm}
\makeatletter
\pdfpageheight\paperheight
\pdfpagewidth\paperwidth
\usepackage[colorlinks,linkcolor=blue,anchorcolor=black,citecolor=red]{hyperref}

\makeatother

\begin{document}

\title[Article Title]{Non-Hermitian unidirectional routing of photonic qubits}

\author{En-Ze Li}
\thanks{These authors contributed equally to this work}
\affiliation{Key Laboratory of Quantum Information, University of Science and Technology of China, Hefei, Anhui 230026, China.}
\affiliation{Synergetic Innovation Center of Quantum Information and Quantum Physics, University of Science and Technology of China, Hefei, Anhui 230026, China.}

\author{Yi-Yang Liu}
\thanks{These authors contributed equally to this work}
\affiliation{School of Physical Science and Technology, Lanzhou University, Lanzhou, Gansu 730000, China.}

\author{Ming-Xin Dong}
\affiliation{Key Laboratory of Quantum Information, University of Science and Technology of China, Hefei, Anhui 230026, China.}
\affiliation{Synergetic Innovation Center of Quantum Information and Quantum Physics, University of Science and Technology of China, Hefei, Anhui 230026, China.}

\author{Dong-Sheng Ding}
\email{dds@ustc.edu.cn}
\affiliation{Key Laboratory of Quantum Information, University of Science and Technology of China, Hefei, Anhui 230026, China.}
\affiliation{Synergetic Innovation Center of Quantum Information and Quantum Physics, University of Science and Technology of China, Hefei, Anhui 230026, China.}
\affiliation{Hefei National Laboratory, Hefei, Anhui, 230088, China}

\author{Bao-Sen Shi}
\email{drshi@ustc.edu.cn}
\affiliation{Key Laboratory of Quantum Information, University of Science and Technology of China, Hefei, Anhui 230026, China.}
\affiliation{Synergetic Innovation Center of Quantum Information and Quantum Physics, University of Science and Technology of China, Hefei, Anhui 230026, China.}
\affiliation{Hefei National Laboratory, Hefei, Anhui, 230088, China}

\begin{abstract}
Efficient and tunable qubit unidirectional routers and spin-wave diodes play an important role in both classical and quantum information processing domains.
Here, we reveal that multi-level neutral cold atoms can mediate both dissipative and coherent couplings.
Interestingly, we investigate and practically implement this paradigm in experiments, successfully synthesizing a system with dual functionality as both a photonic qubit unidirectional router and a spin-wave diode. 
By manipulating the helicity of the field, we can effectively balance the coherence coupling and dissipative channel, thereby ensuring the unidirectional transfer of photonic qubits.
The qubit fidelity exceeds $97.49\pm0.39\%$, and the isolation ratio achieves $16.8\pm0.11$ dB while the insertion loss is lower than 0.36 dB.
Furthermore, we show that the spin-wave diode can effectively achieve unidirectional information transfer by appropriately setting the coherent coupling parameters.
Our work not only provides new ideas for the design of extensive components in quantum network, but also opens up new possibilities for non-Hermitian quantum physics, complex quantum networks, and unidirectional quantum information transfer.
\end{abstract}
\date{\today}
\maketitle

\textit{Introduction.}---In contemporary communication and information technology, the concept of unidirectional routing for information carriers has garnered substantial attention. 
This concept is extensively employed in the transmission of various signals, including acoustic waves \cite{devaux2019acoustic,wang2022acoustic,hackett2023non}, radio frequencies \cite{nagulu2020non,shao2022electrical}, and quantum signals \cite{lodahl2017chiral}. 
Among them, gyrators serve an indispensable role as key components in facilitating efficient and orderly information exchange between different nodes \cite{nagulu2020non,viola2014hall}; 
dual-port isolators effectively suppress reverse noise \cite{jalas2013and,scheucher2016quantum,lodahl2017chiral,nagulu2020non};
while unidirectional amplifiers focus on the directional amplification of weak signals \cite{wanjura2020topological,shen2018reconfigurable}.
In linear systems, the achievement of unidirectional responses hinges on the disruption of time-reversal symmetry through the application of real or synthetic magnetic fields. 
However, the practicality of these traditional unidirectional devices is hampered by their biased magnetic fields.
In recent years, promising physical mechanisms have emerged to overcome the aforementioned limitations, including nonlinear optics \cite{fan2012all,cao2017experimental,xia2018cavity}, optomechanics \cite{manipatruni2009optical,shen2016experimental}, atomic gases \cite{li2020experimental,dong2021all,zhang2018thermal,hu2021noiseless,pucher2022atomic}, quantum dots \cite{antoniadis2022chiral}, and metamaterials \cite{iyer2020unidirectional}.

The Unidirectional router and spin-wave diode simplify the intricate nature of photonic networks \cite{jalas2013and,metelmann2018nonreciprocal}, augment communication channel capacities \cite{miller2010optical,verhagen2017optomechanical}, and becomes valuable resources in quantum sensing \cite{lau2018fundamental}.
Such a device promotes the development of more efficient and adaptable quantum information platforms \cite{kimble2008quantum,lodahl2017chiral}.
It stimulated numerous recent studies on nonreciprocal couplings and chiral magnons transfer, such as quantum transistors and transducers \cite{gorniaczyk2014single,wang2022ultra,stannigel2010optomechanical}, quantum diodes \cite{hamann2018nonreciprocity,muller2017nonreciprocal,barzanjeh2018manipulating}, unidirectional amplifiers \cite{metelmann2015nonreciprocal,lecocq2017nonreciprocal,peterson2017demonstration,malz2018quantum}, and spin-wave diode \cite{cirac1997quantum,kimble2008quantum,lan2015spin,stannigel2010optomechanical,lodahl2017chiral}.
However, as the quantum nodes increase, the cumulative effect of insertion loss and quantum coherence loss leads to a significant increase in the complexity of directional transmission and detection tasks of quantum states between nodes \cite{lodahl2017chiral}.
Therefore, aside from large isolation, ensuring the high efficiency (low insertion loss) and high fidelity of unidirectional routers and spin-wave diodes is an urgent and promising research topic.

\begin{figure*}[t]
\includegraphics[width=1\linewidth]{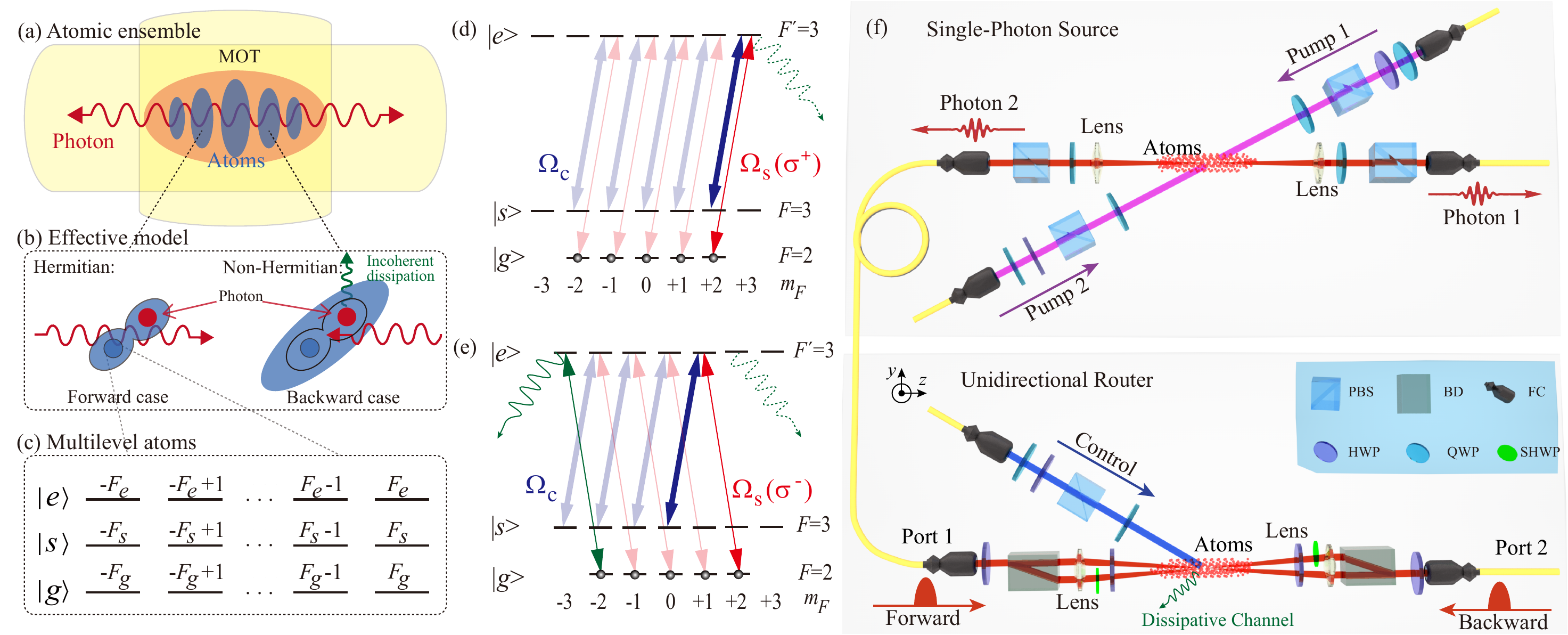}
\caption{\textbf{Schematic and experimental setup of non-Hermitian unidirectional routing}
(a) Sketch of the magnons-photon interface in cold-atoms ensemble.
(b) An efficient non-Hermitian interaction model between light and atoms. In forward routing, coherent coupling produces a photon routing channel. In backward routing, the incoherent coupling leads to dissipation.
(c) Multi-level structure of atoms, which is composed of three states $|g\rangle$, $|s\rangle$, and $|e\rangle$ (magnons) with angular momenta $F_{g}$, $F_{s}$, and $F_{e}$, respectively.
(d) In forward routing, the helicity of control field and signal photon are the same, collective atomic system and photon build coherent coupling.
(e) In backward routing, due to the opposite helicity between the signal photon and the control field, a magnon ($|g,\;m_{F}=-2\rangle$) generates an incoherent dissipative channel.
(f) Schematic diagram of experimental setup.
The single-photon source emits a pair of heralded photons (Photon 1 and 2) \cite{supplimental}. 
The unidirectional router demonstrate the transmission of qubits.
PBS, polarized beam splitter; BD, beam displacer; HWP, half-wave plate; QWP, quarter-wave plate; SHWP, small half-wave plate; FC, fiber collimator.
}
\label{Fig.1}
\end{figure*}

\begin{figure*}[t]
\includegraphics[width=1.0\linewidth]{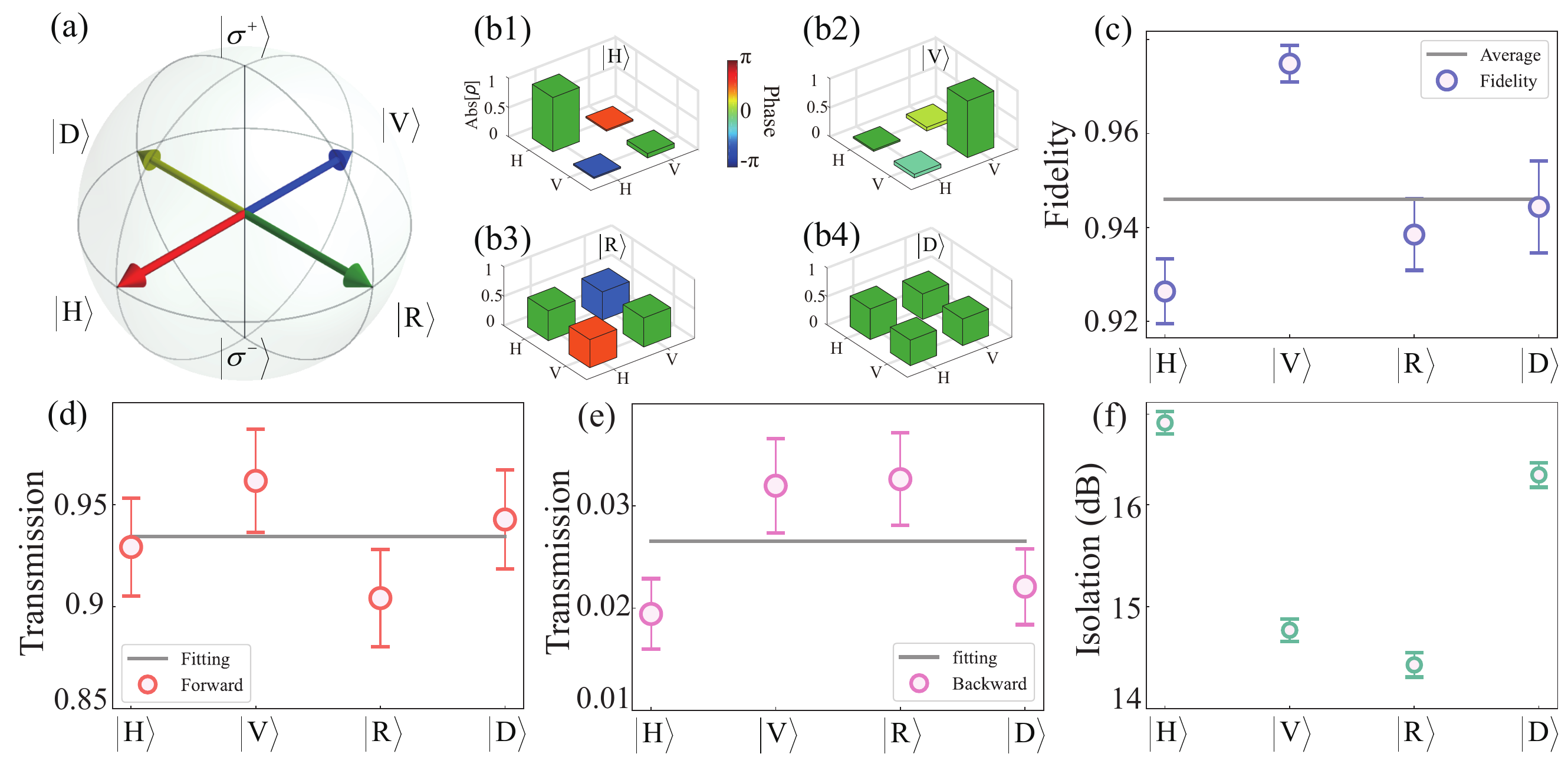}
\caption{
\textbf{Photonic qubits transmission properties.}
(a) The Poincar$\rm \acute{e}$ sphere representation of photonic qubits.
(b1-b4) The reconstructed density matrices of photonic qubits $|{\rm H}\rangle$, $|{\rm V}\rangle$, $|{\rm R}\rangle$ and $|{\rm D}\rangle$ after passing through NHUR along the forward routing.
The height of the bars signifies the absolute value of $\rho$, with the color indicates the phase of the density matrix elements.
(c) The fidelity of the four photonic qubits.
(d) and (e) The routing properties of photonic qubits are observed in the NHUR apparatus, where (d) and (e) correspond to the forward and backward routing, respectively.
The gray lines represent the theoretical fitting.
(f) The isolation ratio of the photonic qubits. 
The error bars are estimated from Poisson statistics and represent ± one standard deviation.
}
\label{Fig.2}
\end{figure*}

\begin{figure*}[t]
\includegraphics[width=1\linewidth]{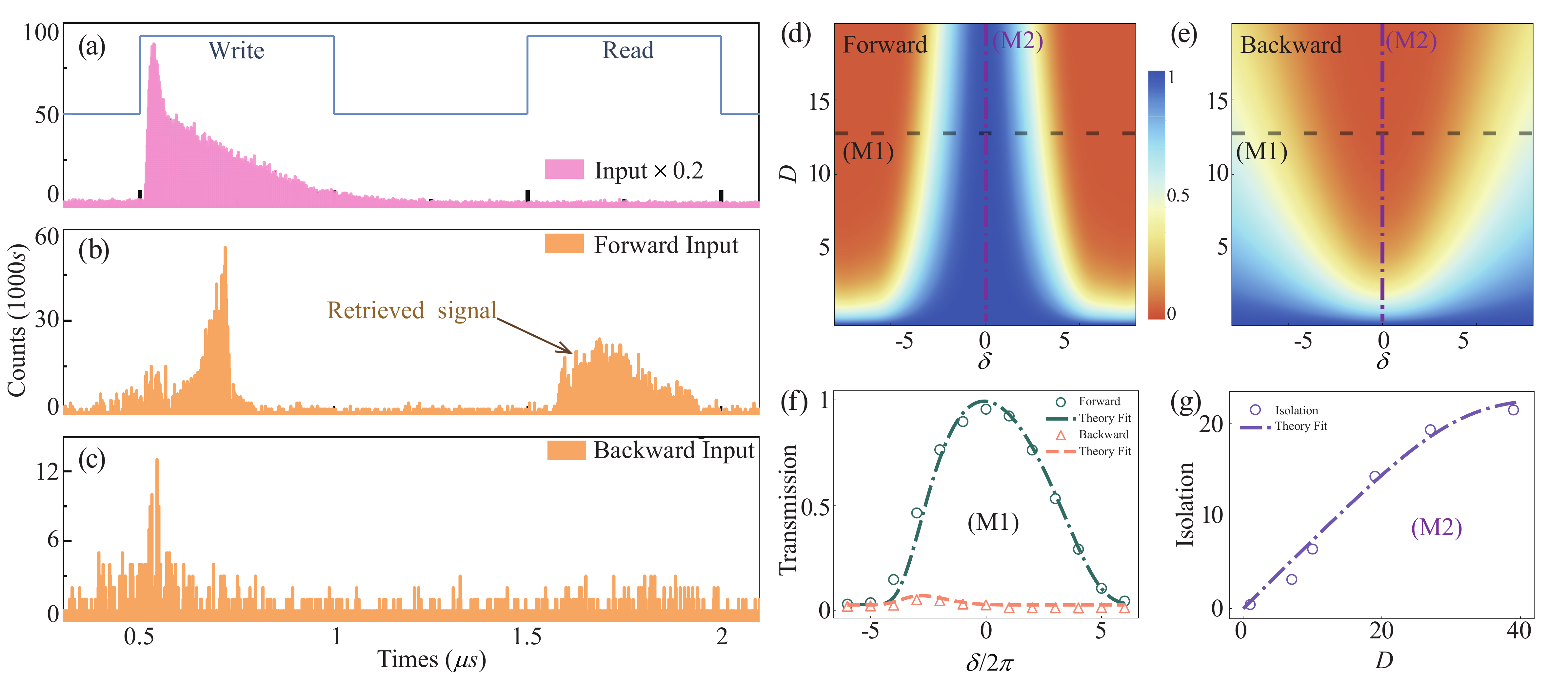}
\caption{
\textbf{Apin-wave diode and dynamical properties of the NHUR apparatus.}
(a) The input wave packet of a photonic qubit prior to the NHUR.
The blue line represents the temporal sequence of the control field.
(b), (c) The retrieved signals of NHUR in forward routing (b) and backward routing (c), respectively.
(d), (e) The output signal spectrum of the forward (d) and backward (e) as a function of optical depth ($D$) and frequency detuning $\delta$.
The dynamical properties of the NHUR experiment, observed at specific parameter configuration locations M1 and M2, are represented by dashed and dotted lines in (f) and (g), respectively.
(f) The transmission spectrum of the forward (green $\Circle$) and backward (orange $\triangle$) routing by scanning the control field detuning $\delta$.
The green dotted line and orange dashed line represent the fitting results obtained at M1 ($D=14$).
(g) The isolation ratio of the NHUR is characterized by the parameter $D$, with the control field detuning $\delta=0$. The dotted line represents a fitting derived from theoretical expectations.
}
\label{Fig.3}
\end{figure*}

\begin{figure}[t]
\includegraphics[width=1\linewidth]{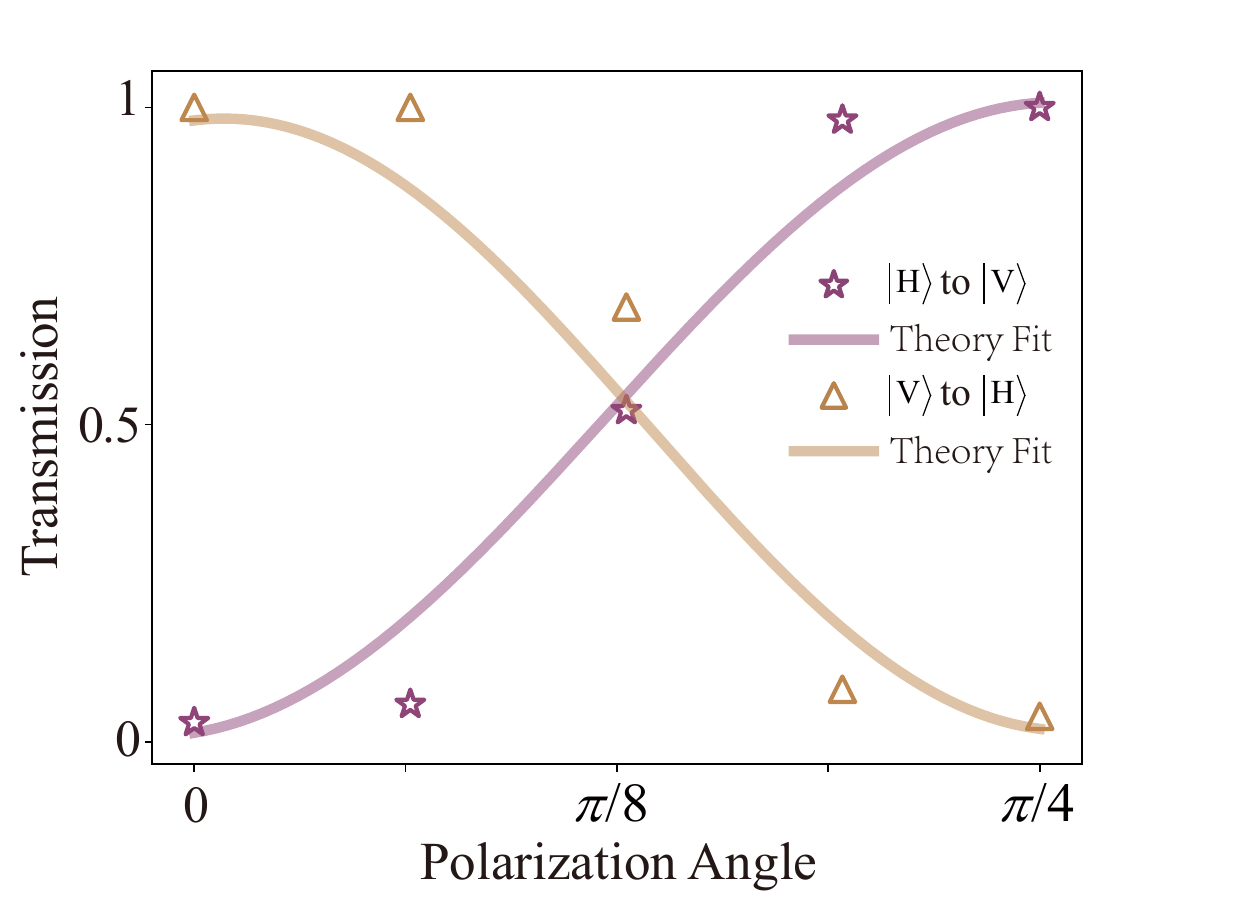}
\caption{
\textbf{Tunable properties of dissipation-induced NHUR.}
The input photonic qubits are initially prepared in $|\rm H\rangle$ (purple star) and $|\rm V\rangle$ (yellow triangle) states, respectively.
The unidirectional transmission characteristics of NHUR can be reversed through by manipulating the chirality of the control field.
Purple stars and yellow triangles correspond to the control field with helicity $\sigma=+1$ and $\sigma=-1$, respectively.
The solid curves represent theoretical fit results.
}
\label{Fig.4}
\end{figure}

In this Letter, we present a paradigm for realizing a non-hermitian unidirectional router (NHUR) and spin-wave diode for arbitrary photonic qubits.
We consider rich multi-level neutral cold atoms as depicted in Fig.~\ref{Fig.1}(a). 
We demonstrate that the effective interactions utilized in this paradigm can be described by a coherent coupling balanced against its corresponding dissipative version.
The interaction dynamics between the collective atomic states and photons are precisely manipulated through a chiral control field. \cite{wen2019non}.
The helicity and frequency detuning of the control field serve as critical controllable parameters that generate non-Hermitian behavior, thereby resulting in unidirectional router and spin-wave diode \cite{simon2007single}.
This framework provides a perspective for investigating quantum state transfer mechanisms within chiral quantum networks and information processing domains, paving the way for further investigations \cite{lodahl2017chiral,ye2023photonic,kimble2008quantum,ding2015raman,cho2010atomic,cirac1997quantum,gonzalez2015chiral,lodahl2017chiral}.

\textit{Model and setup.}---Our model is based on the eigenstate structure and cooperative excitation properties arising from the collective coupling of trapped atoms, as shown in Fig.~\ref{Fig.1}(b).
Driven by chiral control field, magnons, and collective atomic spin waves in atoms generate a helicity-dependent incoherent dissipation channel.
Balanced coherent coupling and dissipative channel break the time-reversal symmetry \cite{gorshkov2007universal,duan2001long,wen2019non}. 
In this scenario, dissipation profoundly alters the routing matrix, transforming it from a Hermitian to a non-Hermitian configuration \cite{jalas2013and}.
This process induces directional transfer properties in the photon-atom interface.

The multi-level atoms are consisted in two metastable states $|g\rangle$ and $|e\rangle$ (magnons), and an excited state $|s\rangle$ \cite{chen2000roles,wang2006controlling}, as shown in Fig.~\ref{Fig.1}(c). 
The underlying directional quantum processes is explained by the energy-level diagram [Fig.~\ref{Fig.1}(d) and Fig.~\ref{Fig.1}(e)].
Initially, all atoms were prepared in the $|g\rangle$ state, where $m_{F}$ represents the magnetic quantum number of the hyperfine states.
A control field ($\Omega_{c}$) drives coherence transition between $|s\rangle$ and $|e\rangle$. 
Signal photons couple the transition between $|g\rangle$ and $|e\rangle$.
The helicity of the control field is fixed to $\sigma=1$, indicating the atomic transition from states $\left|s\right\rangle$ to $\left|e\right\rangle$ with $\Delta m_{F}=+1$ \cite{alpeggiani2018electromagnetic}.
The chirality of the optical field as an effective tuning knob can balance the coherent coupling and dissipative channels.

In forward routing, as depicted in Fig.~\ref{Fig.1}(d), the helicity of the signal photon is $\sigma=1$.
The coherent coupling exists between the signal photon and the collective atomic states, forming a stable photon channel between two states. 
In backward routing, the photon follows the spin-momentum locking \cite{bliokh2014magnetoelectric,van2016universal,yang2020spin}, entailing a time-reversal operation as $\mathbf{k}\rightarrow-\mathbf{k}, \sigma\rightarrow-\sigma$, as shown in Fig.~\ref{Fig.1}(e). 
The chiral photon induces a dissipation channel that couples the state $|g,m_F=-2\rangle$ and $|e,m_F=-3\rangle$, mediating incoherent interactions between the collective exciton and its immediate environmental surroundings \cite{wen2019non}.

In the rotating frame, we obtain an effective non-Hermitian Hamiltonian $H_{\text{eff}}$ governing the dynamics of magnons-photon interface, as the form of \cite{supplimental}
\begin{equation}
H_{\text{eff}}=
\left(
\begin{array}{ccc}
\Lambda     &     0             & \Omega_p/2          \\
0                  &    \delta         & \Omega_c/2          \\
\Omega_p/2         &    \Omega_c/2     & \Delta_p               
\end{array}
\right)
\end{equation}
\noindent where $\delta=\Delta_{p}-\Delta_{c}$ is the two-photon detuning, $\Delta_{c}$ and $\Delta_{p}$ denote the detuning of the control and signal field, respectively.
The parameter $\Lambda$ determines the level of non-Hermiticity in the system (the calculated eigenvalues are complex values when $\Lambda$ is non-zero \cite{supplimental}). 
In forward routing, $\Lambda^{\text{fw}}=0$, $H_{\text{eff}}$ is a Hermitian operator.
In backward routing, $\Lambda^{\text{bw}}=-i\gamma_{\text{eff}}/2$, the $H_{\text{eff}}$ transforms into non-Hermitian Hamiltonian.  
$\gamma_{\text{eff}}=\Omega^2/\Gamma$ arises from the incoherent dissipation, $\Omega$ is the Rabi frequency of the dissipative coupling, and $\Gamma$ is the spontaneous decay rate of the state $|e\rangle$. 

Based on the above model, we design the experimental setup with the schematic diagram shown in Fig.~\ref{Fig.1}(f).
The system consists of two components: a single-photon source and a unidirectional router.
The single-photon source emits heralded Stokes (Photon 1, S1) and anti-Stokes (Photon 2, S2) photons \cite{supplimental,balic2005generation,du2008narrowband}.
The S1 serves as a trigger signal, and the unidirectional router introduces S2 via ports 1 or 2 to investigate routing properties.
Here, we establish the $z$-axis as the quantization axis and designate the terms `forward' and `backward' to illustrate the routing direction of S2 along $+z$ and $-z$ axes, respectively.
We utilize a beam displacer to displace the arbitrary qubit state ($|\psi\rangle ={\rm cos}\frac{\theta}{2}|\rm H\rangle +e^{\emph{i}\phi}{\rm sin}\frac{\theta}{2}|\rm V\rangle $) onto a pair of orthogonal polarized states, $|{\rm H}\rangle$ and $|{\rm V}\rangle$, where they are spatially separated. 
Then, we configure the two photon paths to maintain a same helicity.
After passing through the atoms, we employ the reverse process to erase the path information and reconstruct the qubits.
Meanwhile, the control field is incident on the atoms at $2^{{^\circ}}$ with respect to the signal path.

The time evolution of the density matrix $\rho$ is described by the master equation ${\dot\rho}=-i[H_{\text{eff}},\rho]+{\mathcal{L}}[\rho]$, where $\mathcal{L}$ is the Lindblad operator describing the decay and dephasing of the atoms \cite{supplimental}.
The signal transmission is proportional to $T=|h|^2$ ($h=\mathrm{exp}[-D\widetilde{\chi}/2]$), where $D$ is the optical depth of atoms.
$\widetilde{\chi}=\chi_{\rm EIT}+\chi_{\rm loss}$ is the effective chiral susceptibility of atoms, and the imaginary part of $\Tilde{\chi}$ indicates the absorption \cite{supplimental}.
$\chi_{\text{EIT}}$ illustrates the coherent coupling due to electromagnetically induced transparency (EIT) \cite{duan2001long}.
$\chi_{\text{loss}}$ component accounts for incoherent dissipation which could result in photon losses \cite{supplimental,fleischhauer2005electromagnetically,wen2019non}.

By solving the master equation, we derive the chiral susceptibility within the theoretical framework.
In forward routing, $\chi_{\rm loss}=0$, the effective chiral susceptibility is solely determined by the coherent coupling, taking the form \cite{supplimental}.
\begin{equation}
\widetilde\chi_{1\to2}=\chi_{\text{EIT}}^{1\to2}=\sum\limits_{i=-2}^{2}\eta\;G_{i+1,i}\;\rho_{i+1,i}\;.
\label{eq:2}
\end{equation}
\noindent In backward routing, dissipative channels and coherent coupling compete with each other, and the effective susceptibility is given by
\begin{align}
\widetilde\chi_{2\to1}=&\chi_{\text{EIT}}^{2\to1}+\chi_{\text{loss}}^{2\to1}\\
=&\sum\limits_{i=-2}^{2}\eta\;G_{i-1,i}\;\rho_{i-1,i}+\eta\;G_{-3,-2}\;\rho_{-3,-2}\;,
\label{eq:3}
\end{align}
\noindent where $\eta$ depends on the initial atomic population, $G_{mn}=2N|\mu_{e_{m},g_{n}}|^{2}/(\hbar\varepsilon_{0})$ describes the atoms-photon interaction coefficient, and $\mu_{e_{m},g_{n}}$ is the dipole moment for the transitions of $\ensuremath{|g,m_{F}=m\rangle \to|e,m_{F^{\prime}}=n\rangle}\ (m\in \{-3,3\},\ n\in \{-2,2\})$.

\textit{Unidirectional router.}---The $\rm{Poincar\acute{e}}$ sphere of four photonic states $|i\rangle$ are dipicted in Fig.~\ref{Fig.2}(a), where $i=\{\rm H,V,R,D\}$.
Utilizing quantum state tomography \cite{PhysRevA.64.052312,supplimental}, as depicted in Figs.~\ref{Fig.2}(b), we obtain the reconstructed density matrices for the forward-transmitted qubits.
The fidelity metrics of the four qubits are denoted as $\mathcal{F}_{i}=\{0.92,0.97,0.93,0.94\}$, as illustrated in Fig.~\ref{Fig.2}(c). 

The directional transmission rates $T_{i}$ of the four qubits are showned in Fig.~\ref{Fig.2}(d) and (e).
In forward routing, the coherent coupling leads to a high transmission $T^{1\to2}_i=\{0.93,0.96,0.94,0.94\}$. 
As expected, the backward transmission remains negligibly small, with $T^{2\to1}_i=\{0.019,0.032,0.033,0.022\}$. 
Then, in Fig.~\ref{Fig.2}(f), We calculate the isolation ratios of four qubits \cite{supplimental}, and have $I_i=\{16.80,14.77,14.43,16.29\}~{\rm dB}$.
The insertion loss is defined as $-10{\rm log}(T^{1\to2})$, we calculate the four qubits insertion loss with $\{0.36,0.18,0.27,0.27\}~{\rm dB}$. 
The noisy photons produced by the spontaneous-Raman-scattering process in atoms are the main factor limiting our isolation ratio, and the signal-to-noise ratio of signal photons can be further improved by using the optical cavity \cite{hu2021noiseless}.

\textit{spin-wave diode.}---Here, we further achieve spin-wave diode at the magnons-photon interface by utilizing chiral control pulses (write and readout pulses). 
Driven by the write pulse, photonic qubits ($|\rm H\rangle$) are converted into collective excitation spin-wave excitons through EIT and stored in atoms in the form of magnons, as illustrated in Fig.~\ref{Fig.3}(a). 
After a storage time of $500ns$, the magnons are reconverted into qubits through readout pulses, with both the write and readout pulses having a duration of 1 $\mu s$. 
In forward routing, we obtain the retrieved qubit as illustrated in Fig.~\ref{Fig.3}(b).
In Fig.~\ref{Fig.3}(c), no retrieved qubit can be obtained in the backward due to the incoherent dissipative channel. 
The above-described directional state transfer process can thus as function as a spin-wave diode.
This process filters out incoherent noise while protecting the quantum information from unwanted backaction noise \cite{lodahl2017chiral}.
Spin-wave diodes also have great potential for various areas, including directional state transfer in chiral quantum networks \cite{lodahl2017chiral,ye2023photonic}, design of quantum circuits \cite{kimble2008quantum,ding2015raman,cho2010atomic}, the generation of entanglement \cite{cirac1997quantum,gonzalez2015chiral}, and provide a fertile ground for quantum many-body system and collective effects research \cite{lodahl2017chiral}.

\textit{Tunable dynamic properties.}---
We next investigate the dynamical properties of the unidirectional router.
From the theoretical model, the parameters $\delta/\Gamma$ and $D$ play an important role in balancing the coherent coupling and dissipative channel.
Then, We theoretically calculated the expected unidirectional qubits transfer for the parameters $\delta$ and $D$. 

Figure.~\ref{Fig.3}(d) and Fig.~\ref{Fig.3}(e) depict theoratical calculation results of the forward and backward routing transmission, respectively.
We examine the experimental observations (in Fig.~\ref{Fig.3}(f)) at feature position, M1, as indicated by the dotted line in Fig.~\ref{Fig.3}(d).
With a far-detuned control field, i.e., $\Tilde{\chi}_{1\to2}\approx\Tilde{\chi}_{2\to1}\approx-1/\delta$, the atoms absorbs bidirectional signal photons.
When $\delta=0$, the system breaks time-reversal symmetry, leading to the emergence of extreme values in the direction-dependent transfer coefficients, i.e., $h_{1\to2}\approx 1$ and $h_{2\to1}\approx 0$. 
In the vicinity of $\Delta\omega_{p}\sim0$ MHz, we observe significantly high and nearly negligible transmissions in the forward and backward routing respectively. This results in a high forward (low backward) transmission rate of $92.9\pm2.0\%$ ($2.6\pm0.4\%$). 
Figure \ref{Fig.3}(g) represents the observation result of the dotted line M2. 
Limitied by our experimental setup, the isolaiton ratio of the NHUR can reach 20 dB when $D=40$. Theoretically, the isolation can be improved by applying a higher $D$.
Thus, the NHUR process with increased $D$ can be treated as a sequence of spatially cascaded collective-exciton mode-based NHUR.

In Fig~\ref{Fig.4}, we investigate the tunable properties of the NHUR process with a fixed signal helicity.
We initialized the photonic qubits in two states: $|\rm{H}\rangle$ (represented by purple stars) for the forward direction and $|\rm{V}\rangle$ (represented by yellow triangle) for the backward direction.
By continuously adjusting the helicity of the control field \cite{supplimental}, we can dynamically manipulate the qubit transfer direction.
The inclusion of above feature underpin applications of unidirecitonal router within the realm of quantum-information processing by providing a new angle for realizing universal quantum gates \cite{koshino2010deterministic,duan2004scalable}.

\textit{Conclusion.}---
In summary, we investigated the Hamiltonian describing the photon-atoms interaction, demonstrating that multi-level atoms can mediate coherent and dissipative couplings. 
When coherent coupling and dissipative channels are balanced, we obtain a polarization-independent unidirectional router and spin-wave diode. 
Experimental results show that our approach enables high-fidelity, high-isolation, and low-insertion-loss directional transmission of arbitrary polarized qubits. 
Additionally, the spin-wave diode facilitates directional quantum state transfer. This mechanism can be triggered by light fields, magnetic fields, and acoustic waves, potentially coupling with various energy modes. 
Our theoretical framework is expected to apply to physical systems characterized by rich energy level structures, including various similar systems such as molecules, NV centers, trapped ions, and optical microresonators \cite{PhysRevLett.113.263601,PhysRevLett.113.263602,PhysRevA.97.052315,cao2023probing,cao2015dielectric,el2018non,bernon2013manipulation,zhu2020dielectric}. 
Our work not only provides a new perspective for creating effective non-coherent dissipative channels and promotes effective control of the system Hamiltonian but also paves the way for designing quantum diodes and improving existing unidirectional routers \cite{lodahl2017chiral}.

We thank Prof.~Franco Nori, Prof.~Ke-Yu Xia, Prof.~Wei Yi, Prof.~Wei Zhang, Dr.~Ying-Hao Ye, Dr.~Wei-Hang Zhang, and Dr.~Lei Zeng for the fruitful discussions.
This work is supported by the National Key Research and Development Program of China (2022YFA1404002), the National Natural Science Foundation of China (Grant No. U20A20218, 11934013), the Major Science and Technology Projects in Anhui Province (Grant No. 202203a13010001), the Youth Innovation Promotion Association of Chinese Academy of Sciences through Grant No. 2018490, the Innovation Program for Quantum Science and Technology (2021ZD0301100), and Anhui Initiative in Quantum Information Technologies (AHY020200). 

%

\end{document}